\documentclass{svjour2}                    
\smartqed  
\usepackage{graphicx}
\usepackage{amssymb}
\usepackage{amsmath}
\usepackage{amscd}
\usepackage{mathrsfs}
\usepackage{latexsym}
\usepackage{makeidx}
\usepackage{mathptmx}      
\journalname{Foundations of Physics}
\begin{document}

\title{The Quantum Harmonic Oscillator in the ESR Model
}

\titlerunning{The Quantum Harmonic Oscillator in the ESR Model} 

\author{Sandro Sozzo}

\authorrunning{Sandro Sozzo} 

\institute{
           S. Sozzo \at
           Center Leo Apostel (CLEA), Brussels Free University (VUB), Belgium, EU	\\				
           Department of Mathematics and Physics, University of Salento, Lecce, Italy, EU \\
              \email{ssozzo@vub.ac.be, Sozzo@le.infn.it}           
}

\date{Received: date / Accepted: date}

\maketitle

\begin{abstract}
The \emph{ESR model} proposes a new theoretical perspective which incorporates the mathematical formalism of standard (Hilbert space) quantum mechanics (QM) in a noncontextual framework, reinterpreting quantum probabilities as \emph{conditional on detection} instead of \emph{absolute}. We have provided in some previous papers mathematical representations of the physical entities introduced by the ESR model, namely observables, properties, pure states, proper and improper mixtures, together with rules for calculating conditional and overall probabilities, and for describing transformations of states induced by measurements. We study in this paper the relevant physical case of the quantum harmonic oscillator in our mathematical formalism. We reinterpret the standard quantum rules for probabilities, provide new expressions for absolute probabilities, and show how the standard state transformations must be modified according to the ESR model.

\end{abstract}

\section{Introduction\label{intro}}
A crucial and problematical feature of the standard interpretation of quantum mechanics (QM) is \emph{nonobjectivity} of physical properties, which follows from a series of ``no--go theorems'', the most important of which are the Bell--Kochen--Specker \cite{b66,ks67} and Bell \cite{b64,chsh69} theorems. To be precise, if one adopts a minimal ``realistic'' position, according to which individual examples of physical systems can be produced \cite{blm91,bgl96}, one can supply an operational definition of objectivity by stating that a physical property $E$ (\emph{e.g.}, the value of an observable) is objective for a given state $S$ of a physical system $\Omega$ if for every individual example of $\Omega$ in the state $S$ the result of an ideal measurement of $E$ does not depend on the measurement context. Then, the Bell--Kochen--Specker theorem provides examples of physical systems and states in which there are nonobjective properties (\emph{contextuality} of QM), while the Bell theorem shows that contextuality may occur also at a distance (\emph{nonlocality of QM}), both features supporting the standard assumption in QM that for every state there are physical properties that are not objective.

Nonobjectivity of physical properties has many puzzling consequences. In particular, it entails the \emph{objectification problem} \cite{blm91,bl96}, \emph{i.e.}, the main and unsolved problem of the quantum theory of measurement \cite{blm91,bs96,b98}, hence several known paradoxes (Sch\"{o}dinger's cat, Wigner's friend, etc.). The debate about the above problems and the foundations of QM is still alive, as witnessed by some recent publications on these issues \cite{puseyetal12,fisc12}.

Trying to avoid the problems above, the author has recently published together with another author several papers \cite{ga02,ga03,gp04,g07,s07,g08,s08,gs09,gs08,sg08,gs09b,gs11a,gs11b,gs11c,gs12} in which an ESR (\emph{extended semantic realism}) model is elaborated whose mathematical apparatus embodies the mathematical apparatus of QM but quantum probabilities are reinterpreted as conditional on detection rather than absolute. The ESR model consists of a microscopic and a macroscopic part. The former is a noncontextual (hence local) \emph{hidden variables} theory, according to which physical properties are objective and the no--go theorems do not hold because of the aforesaid reinterpretation of quantum probabilities. The latter can instead be presented as a self--consistent theory, without mentioning the hidden variables, even if the hidden variables are needed if one has to prove objectivity or to justify the assumptions introduced at a macroscopic level. The new theory introduces a distinction between absolute and conditional on detection probabilities that does not occur in QM. Hence the predictions of the ESR model about the results of experiments checking absolute probabilities are generally different from the predictions of QM (even if the difference depends on some parameters, the \emph{detection probabilities}, which may make it so small that it remains unnoticed at the experimental level). On the contrary, the predictions of the ESR model about the results of experiments checking conditional on detection probabilities (as Aspect's, or similar subsequent experiments; see, \emph{e.g.}, \cite{genovese05} and references therein) coincide with the predictions of QM \cite{gs08}. There are however physical situations in which the difference between the two theories may become relevant and one can contrive experiments to check which predictions better fit experimental data \cite{gs08,gs11a,gs11c,gs12}.

The main features of the ESR model can be summarized as follows.

(i) Each generalized observable is represented by a pair, consisting of the standard quantum representation and a (commutative) family of positive operator valued (POV) measures parametrized by the set of all pure states of the physical system that is considered. Moreover, a generalized projection postulate (GPP) rules the transformations of pure states induced by nondestructive idealized measurements \cite{s08,gs09,sg08,gs11c,gs12}.

(ii) The traditional Bell inequalities, modified Bell inequalities and quantum predictions hold together in the ESR model because they refer to different parts of the picture of the physical world supplied by the model \cite{s07,g08,gs08,gs11a,gs11c}.

(iii) Each proper mixture is represented by a family of pairs, each pair consisting of a density operator and a convex combination of detection probabilities, parametrized by the set of all macroscopic properties characterizing the physical system that is considered. Moreover, a generalized L\"{u}ders postulate (GLP) that generalizes GPP rules the general transformations of proper mixtures induced by nondestructive idealized measurements \cite{gs09b,gs11b,gs11c,gs12}.

(iv) Each improper mixture is represented by a single density operator, as in QM \cite{gs12}.

(v) The different representations of proper and improper mixtures avoid some deep interpretative problems that arise in QM. Furthermore, an experiment with proper mixtures can be envisaged in which the predictions of the ESR model are different from the predictions of QM, thus discriminating empirically between the two theories \cite{gs11b,gs12}.

In this paper, we consider a ``case study'', that is, we apply the mathematical formalism put forward in (i)--(iv) to the study of the \emph{quantum harmonic oscillator} in the ESR model. We chose this example for both its simplicity and the range of its applications \cite{cohenetal,ms96,qcomp01}. Hence, after briefly resuming the essentials of the ESR model that are required for our purposes, namely, the mathematical representations of pure states and generalized observables (Sect. \ref{modello}), we derive rules for calculating conditional and overall probabilities for the energy and position observables of the harmonic oscillator. We show that these rules generalize the standard quantum rules in this specific example (Sect. \ref{harmonic}). We finally consider the transformations of pure states induced by idealized measurements of energy and position, and point out that some interesting conclusions and predictions can be attained also in this case. 

We stress that the results obtained in the present paper on the quantum harmonic oscillator refer to \emph{idealized}, or perfectly efficient, measurements. The difference between the predictions of the ESR model and the predictions of standard QM thus depends on the \emph{intrinsic} detection probabilities. Nevertheless, this does not imply that in every experiment improving efficiencies may lead to results contradicting QM. According to the ESR model, there might be indeed two kinds of experiments, those checking absolute probabilities and those checking conditional on detection probabilities. Experiments of the first kind will yield results that might differ from the results predicted by QM. Experiments of the second kind will yield results agreeing with QM.

\section{The ESR model\label{modello}}
In the next sections, we summarize the basics of the ESR model that are needed in the following \cite{ga03,gp04,gs09,gs08,sg08,gs11c,gs12}.

\subsection{Basic notion and fundamental equations\label{basic}}
According to the ESR model, a physical system $\Omega$ is operationally defined by a pair $(\Pi, {\mathscr R})$, with $\Pi$ a set of \emph{preparing devices} and ${\mathscr R}$ a set of \emph{measuring apparatuses}. Every preparing device, when activated, prepares an \emph{individual example} of $\Omega$ (which can be identified with the preparation act itself if one wants to avoid any ontological commitment). Every measuring device, if activated after a preparing device, yields an \emph{outcome}, that we assume to be a real number.

In the theoretical description a physical system $\Omega$ is characterized by a set ${\mathscr U}$ of \emph{physical objects} and a set $\mathscr E$ of \emph{microscopic properties} at a microscopic level, and by a set $\mathscr S$ of \emph{states} and a set ${\mathscr O}_{0}$ of \emph{generalized observables} at a macroscopic level.

Physical objects are operationally interpreted as individual examples of $\Omega$, while microscopic properties are purely theoretical entities (the hidden variables of the model). Every physical object $x \in {\mathscr U}$ is associated with a set of microscopic properties (the microscopic properties \emph{possessed} by $x$) which is called the \emph{microscopic state} of $x$ and also is a theoretical entity.
 
States are operationally interpreted as classes of probabilistically equivalent preparing devices, following standard procedures in the foundations of QM \cite{bc81,l83}. Every device $\pi \in S \in {\mathscr S}$, when constructed and activated, prepares an individual example of $\Omega$, hence a physical object $x$, and one briefly says that ``$x$ is (prepared) in the state $S$''. Analogously, generalized observables are operationally interpreted as classes of probabilistically equivalent \emph{measuring apparatuses}. Every $A_0 \in {\mathscr O}_{0}$ is obtained by considering an observable $A$ in the set ${\mathscr O}$ of all observables of QM and adding a \emph{no--registration outcome} $a_0 \in \Re$, $a_0 \notin \Xi$ to the set $\Xi$ of all possible values of $A$ on the real line $\Re$, so that the set of all possible values of $A_0$ is $\Xi_0=\{ a_0 \} \cup \Xi$.\footnote{Two remarks are important at this stage. Firstly, the choice of $a_0$ is arbitrary, which implies that $A$ can be generalized in different ways. We shall presently see, however, that physical probabilities do not depend on the choice of $a_0$. Secondly, when the spectrum of the self--adjoint operator $\widehat{A}$ representing the quantum observable $A$ coincides with $\Re$, the condition $a_0 \notin \Xi$ cannot be fulfilled. In this special case, one can consider a bijective function $f: \Re \longrightarrow \Re^{*}$, with $\Re^{*}$ a proper Borel subset of $\Re$, and then substitute $A$ with $f(A)$. Alternatively, one can simply place $a_0$ in the nondiscrete part of the spectrum of $\widehat{A}$ because the orthogonal projection operator $P^{\widehat{A}}(X)$ associated with the Borel set $X$ by the spectral decomposition of $\widehat{A}$ reduces to $0$ whenever $X=\{ a_0 \}$.}

The set ${\mathscr F}_{0}$ of all (\emph{macroscopic}) \emph{properties} of $\Omega$ is then defined as follows,
\begin{equation}
{\mathscr F}_{0} \ =  \ \{ (A_0, X) \ | \  A_0 \in {\mathscr O}_{0}, \ X \in \mathbb{B}(\Re) \},
\end{equation}
where $\mathbb{B}(\Re)$ is the $\sigma$--algebra of all Borel subsets of $\Re$. Hence the subset 
\begin{equation}
{\mathscr F} \ =  \ \{ (A_0, X) \ | \  A_0 \in {\mathscr O}_{0}, \ X \in \mathbb{B}(\Re), \ a_0 \notin X \} \subset {\mathscr F}_{0}
\end{equation}
is in one--to--one correspondence with with the set ${\mathscr G}=\{ (A, X) \ | \  A \in {\mathscr O}, \ X \in \mathbb{B}(\Re) \}$ of all properties associated with observables of QM.

A measurement of a property $F=(A_0, X) \in {\mathscr F}_{0}$ on a physical object $x$ in the state $S$ is then described as a \emph{registration} performed by means of a \emph{dichotomic registering device} whose outcomes are denoted by \emph{yes} and \emph{no}. The measurement yields outcome yes/no (equivalently, $x$ \emph{displays}/\emph{does not display} $F$), if and only if the value of $A_0$ belongs/does not belong to $X$.

The connection between the microscopic and the macroscopic part of the ESR model is established by introducing the following assumptions.

(i) A bijective mapping $\varphi: f \in {\mathscr E} \longrightarrow F \in {\mathscr F} \subset {\mathscr F}_{0}$ exists.

(ii) If $s$ is the microscopic state of a physical object $x$, and an \emph{idealized measurement} of a macroscopic property $F=\varphi (f)$ is performed on $x$, then $s$ determines a probability $p_{s}^{d}(F)$ that $x$ be detected, and $x$ displays $F$ if it is detected and $f \in s$, does not display $F$ if it is not detected or $f \notin s$. For the sake of simplicity, we will consider only idealized measurements in the following.

The ESR model is \emph{deterministic} if $p_{s}^{d}(F) \in \{0,1 \}$, \emph{probabilistic} otherwise. In the former case it is necessarily noncontextual, hence physical properties are objective, because the outcome of the measurement of a macroscopic property on a physical object $x$ depends only on the microscopic properties possessed by $x$ and not on the measurement context. In the latter case one can recover noncontextuality by adding further hidden variables which do not correspond to microscopic properties in ${\mathscr F}$ to the microscopic properties in ${\mathscr E}$ \cite{gs08}.
  
By using the connection between the microscopic and the macroscopic part of the ESR model one can show \cite{gs08} that, whenever the property $F=(A_0, X) \in \mathscr F$ (hence $a_0 \notin X$) is measured on a physical object $x$ in the (macroscopic) state $S$, the overall probability $p_{S}^{t}(F)$ that $x$ display $F$ is given by
\begin{equation} \label{formuladipartenza}
p_{S}^{t}(F)= p_{S}^{d}(F)p_{S}(F) \ .
\end{equation}

The symbol $p_{S}^{d}(F)$ in Eq. (\ref{formuladipartenza}) denotes the probability that $x$ be detected whenever it is in the state $S$ (\emph{detection probability}) and $F$ is measured. The value of $p_{S}^{d}(F)$ is not fixed for a given generalized observable $A_0$ because it may depend on $F$, hence on $X$. But the connection of microscopic with macroscopic properties via $\varphi$ implies that $p_{S}^{d}(F)$ depends only on the features of the physical objects in the state $S$, hence it does not occur because of flaws or lack of efficiency of the apparatus measuring $F$. 

The symbol $p_{S}(F)$ in Eq. (\ref{formuladipartenza}) denotes instead the conditional probability that $x$ display $F$ when it is detected.

Eq. (\ref{formuladipartenza}) only applies to properties in ${\mathscr F}$. But if we consider the measurement of a property $F=(A_0, X)\in {\mathscr F}_{0} \setminus {\mathscr F}$ (hence $a_0 \in X$), it is physically reasonable to assume that, for every $S \in {\mathscr S}$,
\begin{equation} \label{f_bar_c}
p_{S}^{t}(F)=1-p_{S}^{t}(F^{c})=1-p_{S}^{d}(F^{c})p_{S}(F^{c}),
\end{equation}
with $F^{c}=(A_0, (\Re \setminus X) \in {\mathscr F}$. Hence we mainly deal with properties in ${\mathscr F}$ in the following. 

The crucial feature of Eq. (\ref{formuladipartenza}) is that it implies that three basic probabilities occur in the ESR model. We cannot supply as yet any theory which allows us to predict the value of $p_{S}^{d}(F)$. We can however consider $p_{S}^{d}(F)$ as an unknown parameter to be determined empirically, and then introduce theoretical assumptions that connect the ESR model with QM, enabling us to provide  mathematical representations of the physical entities introduced in the ESR model together with explicit expressions of $p_{S}^{t}(F)$ and $p_{S}(F)$.

Let us begin with $p_{S}(F)$. The following statement expresses the fundamental assumption of the ESR model.

\vspace{.1cm}
\noindent
\emph{AX. If $S$ is a pure state and $F \in {\mathscr F}$, the probability $p_{S}(F)$ can be evaluated by using the same rules that yield the probability of $F$ in the state $S$ according to QM.}

\vspace{.1cm} 
Assumption AX allows one to recover the basic formalism of QM in the framework of the ESR model but modifies its standard interpretation. Indeed, according to QM, whenever an ensemble ${\mathscr E}_{S}$ of physical objects in a state $S$ is prepared and ideal measurements of a property $F$ are performed, all physical objects in ${\mathscr E}_{S}$ are detected, hence the quantum rules yield the probability that a physical object $x$ display $F$ if $x$ is selected in ${\mathscr E}_{S}$ (\emph{absolute} probability). According to assumption AX, instead, if $S$ is pure, the quantum rules yield the probability that a physical object $x$ display $F$ if idealized measurements of $F$ are performed and $x$ is selected in the subset of all objects of ${\mathscr E}_{S}$ that are detected (\emph{conditional} probability).

For the sake of simplicity, we limit ourselves here to consider pure states only from now on. Proper and improper mixtures have been analysed both from a physical and a mathematical point of view in the ESR model \cite{gs11a,gs11b,gs11c,gs12}, and they brought in conceptually unexpected novelties, as we have seen in Sect. \ref{intro}. However, they are not relevant for our purposes in this paper.

\subsection{Mathematical representations\label{representations}}
Let us resume the mathematical representations of the physical entities introduced in the ESR model. We present them in an axiomatic form, but they can be deduced from the basics of the ESR model resumed in Sec. \ref{basic} \cite{gs09,sg08,gs11b,gs11c,gs12}. 

\emph{Pure states.} Each physical system $\Omega$ is associated with a (separable) complex Hilbert space ${\mathscr H}$. Let ${\mathscr V}$ be the set of all unit vectors of ${\mathscr H}$. Then, each pure state $S$ of $\Omega$ is represented by a unit vector $|\psi\rangle \in {\mathscr V}$, as in standard QM.

\emph{Generalized observables.} Let $A \in {\mathscr O}$ be an observable of QM and let $A_0$ be the generalized observable obtained from $A$ as specified in Sec. \ref{basic}. We denote by $\widehat{A}$ the self--adjoint operator representing $A$ in QM and by $P^{\widehat{A}}$ the projection valued (PV) measure associated with $\widehat{A}$ by the spectral theorem, 
\begin{equation}
P^{\widehat{A}}: X \in \mathbb{B}(\Re) \longmapsto P^{\widehat{A}}(X)= \int_{X} \mathrm{d} P^{\widehat{A}}_{\lambda} \in {\mathscr L}({\mathscr H})
\end{equation}
(where ${\mathscr L}({\mathscr H})$ is the set of all orthogonal projection operators on ${\mathscr H}$). Then, the generalized observable $A_0$ is represented by the pair
\begin{equation}
(\widehat{A},{\mathscr T}^{\widehat{A}}) ,
\end{equation}
where the second element of the pair is a family ${\mathscr T}^{\widehat{A}}=\{ T_{\psi}^{\widehat{A}}\}_{|\psi\rangle \in {\mathscr V}}$ of positive operator valued (POV) measures
\begin{equation} \label{math_rep_gen_obs}
T_{\psi}^{\widehat{A}}: X \in \mathbb{B}(\Re) \longmapsto T_{\psi}^{\widehat{A}}(X) \in {\mathscr B}({\mathscr H})
\end{equation}
(where ${\mathscr B}({\mathscr H})$ is the set of all bounded operators on $\mathscr H$) defined as follows
\begin{equation} \label{POV_lebesgue}
T_{\psi}^{\widehat{A}}(X) = \left \{
\begin{array}{cll} 
\int_{X}{p}_{\psi}^{d}(\widehat{A}, \lambda) \mathrm{d} P^{\widehat{A}}_{\lambda} & & \textrm{if} \ a_0 \notin X  \\
I - \int_{\Re \setminus X}{p}_{\psi}^{d}(\widehat{A}, \lambda) \mathrm{d} P^{\widehat{A}}_{\lambda}  & & \textrm{if} \ a_0 \in X
\end{array}
\right. 
\end{equation}
(where $I$ is the identity operator on ${\mathscr H}$). The real-valued function ${p}_{\psi}^{d}(\widehat{A}, \cdot)$ in Eq. (\ref{POV_lebesgue}) is such that, for every $|\psi\rangle \in \mathscr H$, $\langle\psi| {p}_{\psi}^{d}(\widehat{A}, \lambda)\frac{d P_{\lambda}^{\widehat{A}}}{d \lambda}|\psi\rangle$ is measurable on $\Re$. Moreover, for every $|\psi\rangle \in {\mathscr V}$, the POV measure $T_{\psi}^{\widehat{A}}$ is commutative, that is, for every $X, Y \in \mathbb{B}(\Re), T_{\psi}^{\widehat{A}}(X)T_{\psi}^{\widehat{A}}(Y)=T_{\psi}^{\widehat{A}}(Y)T_{\psi}^{\widehat{A}}(X)$. 

\emph{Properties.} Let $F=(A_0, X) \in {\mathscr F}_{0}$ be a property of $\Omega$. Then, F is represented by the pair $(P^{\widehat{A}}(X), \{ T_{\psi}^{\widehat{A}}(X) \}_{|\psi\rangle\in {\mathscr V}})$. The first element of the pair coincides with the representation in QM of the property $(A, X)$. The second element is instead a family of \emph{effects} defined by Eq. (\ref{POV_lebesgue}).

\emph{Conditional probabilities.} Let $F=(A_0, X) \in {\mathscr F}$. Then, the conditional probability that a physical object $x$ in the pure state $S$ represented by the unit vector $|\psi\rangle \in {\mathscr V}$ display the property $F$ when $x$ is detected is given by 
\begin{equation} \label{prob_X_QM}
p_{S}(F)=\langle\psi|P^{\widehat{A}}(X)|\psi\rangle .
\end{equation}
The probability in Eq. (\ref{prob_X_QM}) obviously coincides with the conditional probability $p_{S}(A_0, X)$ that an idealized measurement of the generalized observable $A_0$ on $x$ in the state $S$ yield an outcome that lies in the Borel set $X$ when $x$ is detected.

\emph{Overall probabilities.} Let $F=(A_0, X) \in {\mathscr F}$. Then, the overall probability that a physical object $x$ in the pure state $S$, represented by the unit vector $|\psi\rangle \in {\mathscr V}$, display the property $F$ is given by 
\begin{equation} \label{prob_X_W}
p_{S}^{t}(F)=\langle\psi|T_{\psi}^{\widehat{A}}(X)|\psi\rangle .
\end{equation}
The probability in Eq. (\ref{prob_X_W}) obviously coincides with the overall probability $p_{S}^{t}(A_0, X)$ that an idealized measurement of the generalized observable $A_0$ on $x$ in the state $S$ yield an outcome that lies in the Borel set $X$.

Summing up, the representation of a pure state in the ESR model coincides with the standard representation of a pure state in QM. The representations of a generalized observable or of a property is provided instead by means of a pair. In both cases, the first element of the pair coincides with a standard representation in QM and is used to evaluate conditional probabilities, while the second element is a family which is used to evaluate overall probabilities. 

\emph{State transformations.} To close up, we recall the \emph{generalized projection postulate} that rules the transformations of pure states induced by  idealized nondestructive measurements.

\vspace{.1cm}
\noindent
\emph{GPP}. Let $S$ be a pure state represented by the unit vector $|\psi\rangle$, and let an idealized nondestructive measurement of a physical property $F=(A_0,X)\in {\mathscr F}_{0}$ be performed on a physical object $x$ in the state $S$. Let the measurement yield the yes outcome. Then, the state $S(F)$ of $x$ after the measurement is a pure state represented by the unit vector
\begin{equation} \label{genpost_dis_psi}
|\psi(F) \rangle=\frac{T_{\psi}^{\widehat{A}}(X)|\psi\rangle}{\sqrt{\langle\psi | T_{\psi}^{\widehat{A} \dag}(X)T_{\psi}^{\widehat{A}}(X) |\psi\rangle}} \ . 
\end{equation}
Let the measurement yield the no outcome. Then, the state $S'(F)$ of $x$ after the measurement is a pure state represented by the unit vector
\begin{equation} \label{projpostulate_psi_no}
|\psi'(F) \rangle=\frac{T_{\psi}^{\widehat{A}}(\Re \setminus X)|\psi\rangle}{\sqrt{\langle\psi | T_{\psi}^{\widehat{A} \dag}(\Re \setminus X)T_{\psi}^{\widehat{A}}(\Re \setminus X) |\psi\rangle}} \ .
\end{equation}

\vspace{.1cm}
\noindent Eqs. (\ref{genpost_dis_psi}) and (\ref{projpostulate_psi_no}) generalize the projection postulate of QM. 

\subsection{Physical predictions\label{predictions}}
As we have anticipated in Sect. \ref{intro}, the ESR model and QM may yield coincident or different predictions, depending on the kind of the physical experiment that is considered. Indeed, Eq. (\ref{formuladipartenza}) together with assumption AX show that the predictions of the ESR model are theoretically different from those of QM as far as experiments checking overall probabilities are concerned. The predictions of the ESR model coincide instead with the predictions of QM in the case of experiments that actually check conditional on detection probabilities, as Aspect's and similar subsequent experiments \cite{gs08}. From a practical point of view, however, the differences between the predictions of the two theories depend on the values of the detection probabilities, and we have as yet no theory which allows us to calculate these probabilities. In this sense, one could say that the ESR model is incomplete. But, one can consider the detection probabilities in Eqs. (\ref{formuladipartenza}) and (\ref{f_bar_c}), or the real valued functions ${p}_{\psi}^{d}(\widehat{A}, \cdot)$ in Eq. (\ref{POV_lebesgue}), as unknown parameters that can be determined experimentally and then inserted into the equations of the ESR model. This is a common procedure in many theories (see, for example, the free parameters in the minimal standard model), and does not prevent the ESR model from offering a new perspective and yielding new predictions. In particular, even if the detection probabilities are not explicitly predicted by the ESR model, assumption AX implies restrictions on them which have physical consequences that can be experimentally checked, as upper bound on the efficiencies of experimental devices \cite{s07,g08,gs08,gs11a,gs11c}. Moreover, a neat distinction is predicted between the experimental results testing proper and improper mixtures \cite{gs12}. From a theoretical point of view has some important implications. Indeed, it shows that contextuality and nonlocality are not unavoidable consequences of the mere formalism of QM, as commonly maintained:\footnote{This does not mean that well-established ``no--go theorems'', as Bell--Kochen--Specker's and Bell's, are wrong. It means instead that their proofs also depend on implicit interpretative assumptions, and that they can be circumvented if these assumptions are changed.} they also depend on the standard interpretation of quantum probabilities as absolute, and can be avoided if this interpretation is modified, as the ESR model does. Moreover, the objectification problem does not occur and quantum probabilities admit an epistemic interpretation. Obviously, there could be cases in which the foregoing parameters are close to 1, hence the difference between the predictions of the ESR model and those of QM is negligible and remains unnoticed. But, we have recently proved that there are also physically relevant situations in which this difference may be significant and can be experimentally checked \cite{gs08,gs11a,gs11c,gs12}. 

Secondly, we stress that, since we consider only idealized, or perfectly efficient, measurements, the detection probabilities, thus the functions ${p}_{\psi}^{d}(\widehat{A}, \cdot)$, depend only on the intrinsic features of the physical objects that are prepared in the state $S$ represented by the unit vector $|\psi\rangle$, hence they do not depend on flaws or lacks of efficiency of the concrete apparatuses measuring these objects. However, one can straightforwardly extend the mathematical representation of the generalized observables in the ESR model if one wants to take into account nonideal measurements of this kind.

\section{The quantum harmonic oscillator in the ESR model\label{harmonic}}
In this section, we apply the formalism reported in Sect. \ref{modello} to the description of an important physical system, namely the one--dimensional harmonic oscillator.

The simplest example of such a system is a particle $P$ of mass $m$ moving in a potential $V(q)=\frac{1}{2}m \omega^2 q^2$, where $\omega$ is an angular frequency and $q$ is a generalized coordinate. But quantum harmonic oscillators appear everywhere, from state solid physics to quantum field theory \cite{ms96} and quantum computation \cite{qcomp01}. We therefore intend to write down here the explicit expressions of conditional on detection probabilities, the overall probabilities and the state transformations provided by the ESR model for this kind of systems. As predicted by the theory, the former probabilities formally coincide with quantum probabilities but bear a different physical interpretation. The latter probabilities and the state transformations are instead formally different from their counterparts in QM, but may be identified with them FAPP (\emph{for all practical purposes}) if some detection probabilities are sufficiently small.

Let us preliminarily recall some elementary properties of the aforesaid system \cite{cohenetal}.

The observable {\it energy} $H$ of the harmonic oscillator is represented by the Hamiltonian operator $\widehat{H}=\frac{\widehat{P}^{2}}{2m}+\frac{1}{2}m \omega^2 \widehat{Q}^2$ in QM, while the observables {\it position} and {\it momentum} are represented by the self--adjoint operators $\widehat{Q}=\int_{\Re} |q\rangle\langle q| \mathrm{d}q$ and $\widehat{P}=\int_{\Re} |p\rangle\langle p| \mathrm{d}p$, respectively. The spectrum of $\hat{H}$ is given by $\Xi=\{ E_{n}=\hbar \omega (n + \frac{1}{2}) \}_{n \in \mathbb{N}_{0}}$, and the eigenvectors of $\hat{H}$ are provided by the recursive formula $|\phi_{n}\rangle=\frac{1}{\sqrt{n!}} (a^{\dag})^{n}|\phi_{0}\rangle$, where $a^{\dag}$ is a ladder operator. In the $\{| q \rangle \}_{q \in \Re}$ representation, we have
\begin{equation}
\phi_{n}(q)=\langle q|\phi_{n}\rangle=\Big (\frac{m \omega}{\pi \hbar}\Big )^{\frac{1}{4}}\frac{1}{\sqrt{2^{n} n!}}e^{-\frac{m \omega}{2 \hbar}q^{2}}H_{n}(\sqrt\frac{m\omega}{\hbar}q),
\end{equation}
where $\{H_{n} (\sqrt{\frac{m\omega}{\hbar}}q) \}_{n \in {\mathbb N}_{0}}$ are the Hermite polynomials.

Let us now come to the description of the quantum harmonic oscillator in the ESR model. Here the observable {\it energy} $H$ is replaced by a {\it generalized energy} $H_{0}$, in which a no--registration outcome $h_0$ is added to the eigenvalues $E_0, E_1, \ldots, E_n, \ldots$ of the Hamiltonian operator $\widehat{H}$. The generalized energy $H_{0}$ is represented by the pair $(\widehat{H}, \{ T_{\psi}^{\widehat{H}} \}_{\psi \in {\mathscr V}})$, where, for every $|\psi\rangle \in {\mathscr V}$, the POV measure $T_{\psi}^{\widehat{H}}$ is defined by
\begin{equation} \label{energy}
T_{\psi}^{\widehat{H}}(X) = \left \{
\begin{array}{cll} 
\sum_{n, E_n \in X }p_{\psi}^{d}(\widehat{H}, E_n) |\phi_n\rangle\langle\phi_n| & & \textrm{if} \ h_0 \notin X  \\
I -\sum_{n, E_n \in \Re \setminus X} p_{\psi}^{d}(\widehat{H}, E_n) |\phi_n\rangle\langle\phi_n|& & \textrm{if} \ h_0 \in X
\end{array}
\right. 
\end{equation}
where, of course, $p_{\psi}^{d}(\widehat{H},E_n)= p_{S}^{d}((H, \{ E_n \} ))$.

The conditional on detection probability $p_{S}(H_{0}, E_n)$ that an idealized measurement of the generalized energy $H_{0}$ on the particle $P$ in the pure state $S$ represented by the unit vector $|\psi\rangle$ yield the outcome $E_n$ when $P$ is detected coincides with the conditional on detection probability $p_{S}((H_0,\{ E_n\}))$ that a measurement of the property $(H_0, \{ E_n\})$ yield outcome \emph{yes}. Because of assumption AX we obtain
\begin{equation} \label{conditional_energy}
p_{S}(H_0,E_n)=\langle\psi|\phi_n\rangle\langle\phi_n|\psi\rangle=|\langle\phi_n|\psi\rangle|^{2},
\end{equation}
which coincides with the standard quantum formula for probability. Analogously, the overall probability $p_{S}^{t}(H_0,E_n)$ coincides with $p_{S}^{t}((H_{0}, \{E_{n}\}))$ and is given by
\begin{equation} \label{overallenergy}
p_{S}^{t}(H_0,E_n)=p_{\psi}^{d}(\widehat{H}, E_{n})|\langle\phi_n|\psi\rangle|^{2}.
\end{equation}
Furthermore, the overall probability $p_{S}^{t}(H_0,h_0)$ that $P$ be not detected by an idealized measurement of the generalized observable $H_0$ coincides with $p_{S}^{t}((H_0, \{ h_0 \}))$ and is given by
\begin{equation}
p_{S}^{t}(H_0,h_0)=\sum_{n \in {\mathbb N}_{0}} (1-p_{\psi}^{d}(\widehat{H},E_{n})) |\langle\phi_n|\psi\rangle|^{2}.
\end{equation}

By comparing Eqs. (\ref{conditional_energy}) and (\ref{overallenergy}) one gets that they can be identified FAPP whenever the measurement procedure is not idealized and introduces an efficiency $e < p_{\psi}^{d}(\widehat{H},E_{n})$. Similarly, $p_{S}^{t}(H_0,h_0)$ could be indistinguishable from lack of efficiency in real measurement procedures.

Let us come to the formulas that describe the state transformations. Let an idealized measurement of the generalized energy $H_0$ on the particle $P$ in the pure state $S$ represented by the unit vector $|\psi\rangle$ give the outcome $E_n$. Such a measurement is equivalent to a measurement of the property $(H_0, \{ E_n \})$ which yields answer \emph{yes}. Hence, by applying GPP in Sect. \ref{modello}, Eq. (\ref{genpost_dis_psi}), the final state of $P$ after the measurement is a pure state $S_n=S((H_0, \{ E_n \}))$ represented by the unit vector
\begin{equation}
|\psi(( H_{0}, \{ E_n \}))\rangle=|\psi_n\rangle=e^{i \theta_n} |\phi_n \rangle .
\end{equation}
This vector coincides with the one obtained by applying the projection postulate of QM. If the particle $P$ in the pure state $S$ is instead not detected by the idealized measurement of $H_0$, then the final state of $P$ is a pure state $S_{0}=S((H_0, \{ h_0\}))$ represented by the unit vector 
\begin{equation} \label{stateenergynodetection}
|\psi((H_0, \{ h_0 \}))\rangle=|\psi_{0}\rangle=\frac{\sum_{n \in {\mathbb N}_{0}} (1-p_{\psi}^{d}(\widehat{H}, E_{n}))\langle\phi_n|\psi\rangle|\phi_n\rangle}{\sqrt{\sum_{n \in {\mathbb N}_{0}} (1-p_{\psi}^{d}(\widehat{H},E_{n}))^{2} | \langle \phi_n|\psi\rangle|^{2} }} .
\end{equation}

Let us now consider the position observable $Q$ of QM. This observable is replaced in the ESR model by a {\it generalized position} $Q_{0}$, in which a no--registration outcome $q_0$ is added to the spectrum $\Re$ of the position operator $\widehat{Q}$. The generalized position $Q_{0}$ is represented by the pair $(\widehat{Q}, \{ T_{\psi}^{\widehat{Q}} \}_{\psi \in {\mathscr V}})$, where, for every $|\psi\rangle \in {\mathscr V}$, the POV measure $T_{\psi}^{\widehat{Q}}$ is defined by
\begin{equation} \label{position}
T_{\psi}^{\widehat{Q}}(X) = \left \{
\begin{array}{cll} 
\int_{X}{p}_{\psi}^{d}(\widehat{Q}, q) |q\rangle\langle q| \mathrm{d} q  & & \textrm{if} \ q_0 \notin X  \\
I - \int_{\Re \setminus X}{p}_{\psi}^{d}(\widehat{Q}, q) | q\rangle\langle q |\mathrm{d} q  & & \textrm{if} \ q_0 \in X
\end{array}
\right. .
\end{equation}

The conditional on detection probability $p_{S}(Q_0,X)$ that an idealized measurement of the generalized position $Q_{0}$ on the particle $P$ in the pure state $S$ represented by the unit vector $|\psi\rangle$ yield an outcome that lies in the Borel set $X$ when $P$ is detected coincides with the conditional on detection probability that a measurement of the property $(Q_0,X)$ yield the outcome \emph{yes}. Because of assumption AX we obtain
\begin{equation} \label{conditional_position}
p_{S}(Q_0,X)=\langle\psi|P^{\widehat{Q}}(X)|\psi\rangle=\int_{X} |\psi(q)|^{2} \mathrm{d} q ,
\end{equation}
which coincides with the standard quantum formula for probability. Analogously, the overall probability $p_{S}^{t}(Q_0,X)$ is given by
\begin{equation} \label{overallposition1}
p_{S}^{t}(Q_0,X)=\int_{X} {p}_{\psi}^{d}(\widehat{Q}, q) |\psi(q)|^{2} \mathrm{d} q
\end{equation}
if $q_0 \notin X$, while it is given by
\begin{equation}\label{overallposition2}
p_{S}^{t}(Q_0,X)=1-\int_{\Re \setminus X} {p}_{\psi}^{d}(\widehat{Q}, q) |\psi(q)|^{2} \mathrm{d} q
\end{equation}
if $q_0 \in X$.

Finally, let us show that also the formulas for the state transformations are generally different in the ESR model. Indeed, let $q_0 \in X$ and let an idealized nondestructive measurement of the property $(Q_0,X)$ be performed on the particle $P$ in the pure state $S$ represented by the unit vector $|\psi\rangle$ which yields the outcome \emph{yes}. Then, by applying GPP in Sect. \ref{modello}, Eq. (\ref{genpost_dis_psi}) one gets that the final state of $P$ after the measurement is a pure state $S((Q_0,X))$ represented by the unit vector
\begin{equation}
|\psi(X)\rangle=\frac{\int_{X} p_{\psi}^{d}(\widehat{Q},q)\psi(q)|q\rangle \mathrm{d}q}{\sqrt{\int_{X} (p_{\psi}^{d}(\widehat{Q},q))^{2}|\psi(q)|^{2} \mathrm{d}q}} .
\end{equation}
If the particle $P$ in the pure state $S$ is instead not detected by the idealized measurement of $Q_0$, the final state of $P$ is a pure state $T_{0}=S((Q_0, \{ q_0 \}))$ represented by the unit vector 
\begin{equation} \label{statepositionnodetection}
|\psi((Q_0, \{ q_0 \}))\rangle=|\chi_{0}\rangle=\frac{\int_{\Re} (1-p_{\psi}^{d}(\widehat{Q},q))\psi(q)|q\rangle \mathrm{d}q}{\int_{\Re} (1-p_{\psi}^{d}(\widehat{Q},q))^{2}|\psi(q)|^{2}|q\rangle \mathrm{d}q}
\end{equation}
By comparing the vectors in Eqs. (\ref{stateenergynodetection}) and (\ref{statepositionnodetection}), we see that they are generally different. This result shows that, if the particle $P$ is not detected by an idealized nondestructive measurement of a given observable, then its final state generally depends on the observable, and may be different if a different observable is considered.

For the sake of completeness, we also report the expectation values of the generalized observable $H_0$ and $Q_0$ in the pure state $S$ represented by the unit vector $|\psi\rangle$, and compare them with the expectation values of the observables $H$ and $Q$, respectively, in the same state. By assuming $h_0 \in \Re$ and $q_0 \in \Re$, We get
\begin{equation}
\langle H_0 \rangle_{\psi}=h_0 p_{S}^{t}(H_0,h_0)+\sum_{n \in {\mathbb N}_{0}}E_n p_{S}^{t}(H_0,E_n)=h_0+\sum_{n \in {\mathbb N}_{0}}(E_n-h_0)p_{\psi n}^{d}(\widehat{H})|\langle\phi_n|\psi\rangle|^{2} \label{h0}
\end{equation}
\begin{eqnarray}
\langle Q_0 \rangle_{\psi}= q_0 p_{S}^{t}(Q_0,q_0)+\int_{\Re} q p_{\psi}^{d}(\widehat{Q}, q)|\psi(q)|^{2} \mathrm{d} q= \nonumber \\q_0+\int_{\Re} (q-q_0)p_{\psi}^{d}(\widehat{Q}, q)|\psi(q)|^{2} \mathrm{d} q
\label{q0}
\end{eqnarray}
\begin{eqnarray}
\langle H \rangle_{\psi}&=&\langle\psi|\widehat{H}|\psi\rangle=\sum_{n \in {\mathbb N}_{0}} E_n |\langle\phi_n|\psi\rangle|^{2} \label{h} \\
\langle Q \rangle_{\psi}&=&\langle\psi|\widehat{Q}|\psi\rangle=\int_{\Re} q |\psi(q)|^{2} \mathrm{d}q \label{q}
\end{eqnarray}
As expected, Eqs. (\ref{h0}) and (\ref{q0}) are generally different from Eqs. (\ref{h}) and (\ref{q}), respectively. However, if one puts $h_0=0$ and $q_0=0$ (see footnote 1) one gets
\begin{eqnarray}
\langle H \rangle_{\psi}-\langle H_{0} \rangle_{\psi}=\sum_{n \in {\mathbb N}_{0}} E_n (1-p_{\psi}^{d}(\widehat{H},E_{n})) |\langle\phi_n|\psi\rangle|^{2} \label{30}\\
\langle Q \rangle_{\psi}-\langle Q_{0} \rangle_{\psi}=\int_{\Re} (1-p_{\psi}^{d}(\widehat{Q}, q))|\psi(q)|^{2} \mathrm{d}q \label{31}
\end{eqnarray}
which exhibits the relationships between the expectation values considered above.

We conclude this section with a final comment on the results achieved on the quantum harmonic oscillator. As we have observed in Sect. \ref{predictions}, the possibility of discriminating between the ESR and standard QM representations of the observables \emph{energy} and \emph{position} of the quantum harmonic oscillator crucially depends on the numerical values of the detection probabilities $p_{\psi}^{d}(\widehat{H},E_{n})$ and $p_{\psi}^{d}(\widehat{Q}, q)$, respectively. These parameters are not predicted by the ESR model. The only conditions required by the model is that they are such that the conditional on detection expectation values coincide with standard quantum expectation values. But, the detection probabilities can be determined experimentally and then inserted into Eqs. (\ref{30}) and (\ref{31}). If one is now able to perform measurements that are close to ideality, hence if one limits the effects due to flaws or lacks of efficiency, one can then determine these intrinsic parameters by counting the number of physical objects that are prepared in the state $S$ represented by the unit vector $|\psi\rangle$ and the number of physical objects that are not detected by the corresponding measurements. Of course, this procedure may be experimentally difficult. Notwithstanding this, we have recently proved that new kind of experimental check can be envisaged for specific physical examples, namely, pairs of spin--$1/2$ quantum particles in the singlet spin state \cite{gs08,gs11a,gs11c} and ensembles of spin--$1/2$ quantum particles in proper mixtures \cite{gs12}. Similar reasonings apply to the quantum harmonic oscillator, and we plan to discuss this issue in a forthcoming paper.

\section*{Acknowledgement}
The author is greatly indebted with Claudio Garola for reading the manuscript and providing valuable remarks and suggestions.

\end{document}